\documentclass[aps,pre,reprint,superscriptaddress,longbibliography]{revtex4-1}

\usepackage{graphicx}
\usepackage{dcolumn}
\usepackage{bm}
\usepackage{amsmath}
\usepackage{amssymb}
\usepackage{latexsym}
\usepackage{amsbsy}
\usepackage{array}
\usepackage{amssymb}
\usepackage{setspace}
\usepackage{bm}
\usepackage{xcolor}
\usepackage{booktabs}
\usepackage{subfigure}
\usepackage{multirow}
\usepackage[colorlinks,linkcolor=blue,citecolor=blue]{hyperref}
\usepackage{float}

\renewcommand{\vec}[1]{{\ensuremath{\bm{\mathrm{#1}}}}}
\usepackage{dsfont}
\newcommand{\mat}[1]{{\ensuremath{\mathds{#1}}}}

\begin{document}

\title{First-Principles Study of Exchange Interactions of Yttrium Iron Garnet}

\author{Li-Shan Xie}
\affiliation{The Center for Advanced Quantum Studies and Department of Physics, Beijing Normal University, Beijing, 100875, China}
\author{Guang-Xi Jin}
\affiliation{Key Laboratory of Quantum Information, University of Science and Technology of China, Hefei, 230026, China}
\author{Lixin He}
\affiliation{Key Laboratory of Quantum Information, University of Science and Technology of China, Hefei, 230026, China}
\author{Gerrit E. W. Bauer} 
\affiliation{Institute for Materials Research, Tohoku University, Sendai 980-8577, Japan}
\affiliation{WPI Advanced Institute for Materials Research, Tohoku University, Sendai 980-8577, Japan}
\affiliation{Kavli Institute of NanoScience, Delft University of Technology, 2628 CJ Delft, The Netherlands}
\author{Joseph Barker}
\affiliation{Institute for Materials Research, Tohoku University, Sendai
980-8577, Japan}
\author{Ke Xia}
\affiliation{The Center for Advanced Quantum Studies and Department of Physics, Beijing Normal University, Beijing, 100875, China}

\date{\today}

\begin{abstract}
Yttrium Iron Garnet (YIG) is the ubiquitous magnetic insulator used for  studying pure spin currents. The exchange constants reported in the literature vary considerably between different experiments and fitting procedures. Here we calculate them from first-principles. The local Coulomb correction ($U-J$) of density functional theory is chosen such that the parameterized spin model reproduces the experimental Curie temperature and a large electronic band gap, ensuring an insulating phase. The magnon spectrum calculated with our parameters agrees reasonably well with that measured by neutron scattering. A residual disagreement about the frequencies of optical modes indicates the limits of the present methodology.
\end{abstract}

\pacs{}

\maketitle

\section{Introduction}
Yttrium iron garnet (Y$_{3}$Fe$_{5}$O$_{12}$-YIG) is a ferrimagnetic insulator of particular significance due to its uniquely low magnetic damping and relatively high Curie temperature ($\sim570$~K). There has been a recent resurgence in interest after Kajiwara et al. \cite{Kajiwara2010} electrically injected spin waves into YIG  and detected (by the inverse spin Hall effect) their transmission over macroscopic distances of 1 mm. Short wave length spin waves excited electrically \cite{Cornelissen2015}
or thermally \cite{Brandon2015} can also diffuse over distances of 40 $\mu$m,
even at room temperature, demonstrating the potential of using spin waves as information carriers in spintronic applications. The spin Seebeck effect (SSE) in YIG \cite{Uchida2010,Uchida2010apl} also garners attention in the field known as spin caloritronics \cite{Bauer2012}. Recent results on the SSE in the related garnet Gadolinium-Iron Garnet (GdIG)~\cite{Geprags2016} illustrate the importance of understanding the many mode  spin wave spectrum \cite{Xiao:2010uc}.

Most experiments on YIG are interpreted in terms of a single magnon band with parabolic dispersion and a single exchange or spin wave stiffness parameter. However, the magnetic primitive cell contains 20 Fe moments and gives a complicated spin wave spectrum with many modes in the THz range~\cite{Harris1963}. The quantitative quality of  Heisenberg spin models of YIG~\cite{Barker:2016tr} relies on the accuracy of the derived parameters, such as exchange constants and magnetic moments. Through several decades of literature there is a plethora of suggested exchange constants for YIG. All are deduced either from macroscopic measurements such as calorimetry, or are fitted to the neutron scattering data by Plant from 1977~\cite{Plant1977}. The triple axis inelastic neutron scattering only resolved  3 of the 20 spin wave branches which has led to a  quite a spread in exchange parameter. The limited experimental data is insufficient to uniquely fit the exchange parameters. Moreover, the spin wave spectrum of YIG is anomalously sensitive to small changes in the exchange constants. Small changes in the exchange parameters appear to give dramatically different spectra. Here we employ computational material science to improve this unsatisfactory situation.

Different ab initio techniques can be employed to deduce Heisenberg exchange parameters. Within density functional theory (DFT) the Heisenberg Hamiltonian can be fitted to the calculated total energy to identify the coupling constants. There are two common methods of doing this. In the `real-space' method, the total energy of a set of collinear spin configurations with spin flips on different sites is mapped onto the Hamiltonian \cite{Wang2008,Gao2013}. The alternative method is to compute the spin wave stiffness  from the total energy of spin spirals by varying the pitch \cite{Essenberger2011}. For simple, one component systems such as Fe, Co, Ni, both approaches give a good agreement between them selves and also with experimental data~\cite{Halilov1998,Pajda2001}. Here we have chosen to use the real-space method with collinear spin configurations due to the  simplicity of implementation when treating the complex crystal structure of YIG.

\begin{figure}
                \centering

\includegraphics[width=0.4\textwidth]{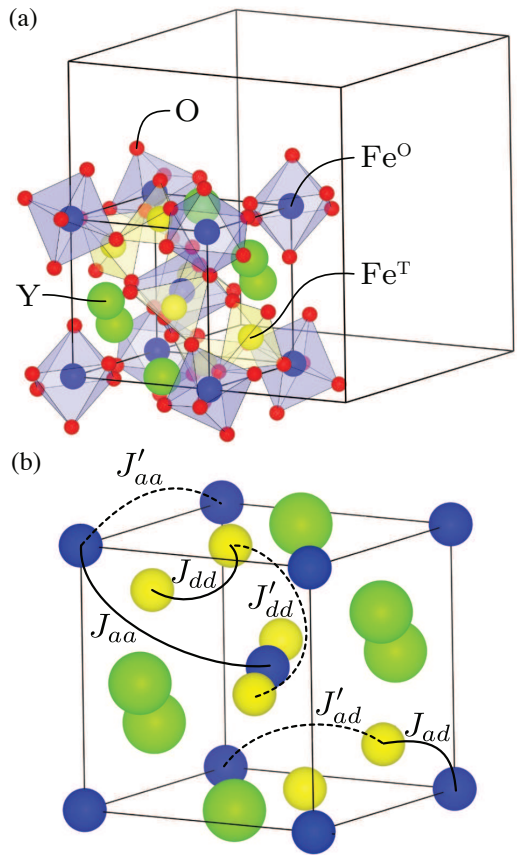}        
        
        \caption{(a) 1/8 of the YIG unit cell. The dodecahedrally coordinated
Y ions (green) occupy the 24c Wyckoff sites, the octahedrally
coordinated Fe$^{\rm O}$ ions (blue) occupy the 16a sites, and the
tetrahedrally coordinated Fe$^{\rm T}$ ions (yellow) occupy the
24d sites. The oxygen (red) 96h sites are not confined by symmetry,
while all cation sites are on special crystallographic positions. (b) The solid and dash lines denote
the nearest and next nearest neighbor exchange interactions. The subscripts $aa$, $dd$ and
$ad$ are stand for the Fe$^{\rm O}$-Fe$^{\rm O}$, Fe$^{\rm T}$-Fe$^{\rm T}$
and Fe$^{\rm O}$-Fe$^{\rm T}$ interactions, respectively. \label{fig1}}
\end{figure}

YIG belongs to the cubic centrosymmetric space group $Ia\overline{3}d$~\cite{Geller1957,Geller1959}. The primitive BCC unit cell contains 80 atoms. One eighth of it is shown in Fig.~\ref{fig1}(a). The magnetic structure as determined by neutron diffraction measurements~\cite{Bertaut1956} confirms that the spins of the Fe$^{\rm O}$ and Fe$^{\rm T}$ ions are locked into an anti-parallel configuration. There is a net magnetization because of the 3:2 ratio of Fe$^{\rm O}$ to Fe$^{\rm T}$ sites in the unit cell, hence YIG is a ferrimagnet.

As a magnetically soft insulator, the magnetism in YIG can be well described by the Heisenberg model
\begin{eqnarray}
E_{tot}&=&E_0-\frac{1}{2}\sum_{i \neq j}J_{ij}\mathbf{S}_i \cdot \mathbf{S}_j ,
\label{eq:energy}
\end{eqnarray}
where $E_0$ is the total energy excluding spin-spin interactions and $\mathbf{S}_i$ is a classical spin vector (of unit length) of the $i$th Fe atom. The exchange interaction $J_{ij}$ is usually considered to be short ranged, but in principle the index is summed over all spins in the crystal.  We initially consider only nearest neighbor (NN) exchange interactions (as done by most previous works); hence there are three independent exchange constants, $J_{aa}$, $J_{dd}$, $J_{ad}$ covering inter- and intra-sublattice interactions as indicated in Fig.~\ref{fig1}(b). Comparing
the energy of the model Hamiltonian (\ref{eq:energy}) with the total energy
calculated ab initio for different spin configurations which should be degenerate in energy, we find unacceptably large energy differences ($\sim2$ meV)
when only including NN interactions. Therefore, later in this work we extend the model to include also next nearest neighbor (NNN) exchange interactions parameterized by three more exchange constants $J'_{aa}$, $J'_{dd}$ and $J'_{ad}$ (also shown in Fig.~\ref{fig1}(b)). Previous works which have included interactions beyond NN~\cite{Plant1983} suffer from an increased over-parameterization of the fitting of only 3 spin wave modes in the neutron scattering data. Our minimal reliance on experimental data puts the justification for the inclusion of NNN on a more solid footing.  

We disregard the magnetocrystalline anisotropy energy which for pure YIG is known to be small and in fact is beyond the accuracy of our methods. The dipolar interactions do not interfere with the exchange energy and can be added a posteriori. The exchange constants are fitted to a number of different collinear spin configurations in which spins are flipped from the ground state. The number of different configurations must be larger than the number of adjustable parameters (3 for the NN model and 6 for the NNN model). 

\section{Exchange fitting}
We now give a brief outline of how the Heisenberg Hamiltonian is mapped onto the different spin configurations. We consider a spin wave of wave vector $\mathbf{k}$ that induces small oscillations in a spin moment $\mathbf{S}_i$ on site $i$ about the collinear ground state.
\begin{eqnarray}
        \phi_i^\mathbf{k}(t)=\mathbf{k}\cdot\mathbf{R}_i+\phi_\alpha(t).
        \label{phi}
\end{eqnarray}
The total energy Eq.~(\ref{eq:energy}) becomes
\begin{eqnarray}
        \begin{aligned}
                E_{ij}^{\phi}(\mathbf{k},\theta,t)= & E_0-\frac{1}{2}\sum_{i \neq j}J_{ij}S_iS_j[\cos\theta_i\cos\theta_j\\
                &+\sin\theta_i\sin\theta_j\cos(\phi_i^\mathbf{k}(t)-\phi_j^\mathbf{k}(t))].
        \end{aligned}
        \label{etotinss}
\end{eqnarray}
The equation of motion for the spin magnetic moments is
\begin{equation}
        \frac{d\mathbf{S}_i(t)}{dt} = -\mathbf{S}_{i}(t)\times\mathbf{H}_{i}
\end{equation}
where $\vec{H}_{i} = -\partial E / \partial \vec{S}_{i}$ is the effective magnetic field. Then 
\begin{eqnarray}
        \begin{aligned}
                \frac{d\phi_j}{dt}\sin\theta_j&=\sum_{i(\neq j)} J_{ij}S_{i}[\cos\theta_i\sin\theta_j\\
                &-\cos(\phi_i^\mathbf{k}-\phi_j^\mathbf{k})\sin\theta_i\cos\theta_j],
        \end{aligned}
        \label{rmeq1}
\end{eqnarray}
If $\theta_i \ll 1$ or $(\pi-\theta_i) \ll1 $, $d\phi/ dt\approx\omega_{\mathbf{k}}$. Expanding Eq.~(\ref{rmeq1}) to lowest order leads to
\begin{equation}
\omega_\mathbf{k}\theta_j=\sum_{i(\neq j)} J_{ij}S_{i}[A_i\theta_j-\cos(\mathbf{k}\cdot\mathbf{d}_{ij})\theta_i A_j],
\label{eq:rmeq2}
\end{equation}
where $\mathbf{d}_{ij}=\mathbf{R}_{i}-\mathbf{R}_{j}$, and the prefactor $A_i$ is +1 for $\theta_i \approx 0$ and -1 for $\theta_i \approx \pi$. The frequencies of the normal modes of this spin system are the eigenvalues
of the matrix $\mat{M}$,
\begin{eqnarray}
\begin{aligned}
\mat{M}_{\alpha\beta}=&\left(\sum_{\gamma}J_{\alpha\gamma}(\mathbf{0})S_{\gamma}A_{\gamma}\right)\delta_{\alpha\beta}-J_{\alpha\beta}(\mathbf{k})S_{\beta}A_{\alpha},
\end{aligned}
\label{rmat}
\end{eqnarray}
\begin{eqnarray}
J_{\alpha\beta}(\mathbf{k})&=&\sum_{d}J_{\alpha\beta} \cos(\mathbf{k}\cdot\mathbf{d}_{ij}) ,
\label{rmat1}
\end{eqnarray}
where the indices $\alpha$ and $\beta$ label the 20 different positions in the unit cell, $\delta_{\alpha\beta}$ is the Kronecker delta, $\mathbf{d}_{ij}=\mathbf{R}_{i}-\mathbf{R}_{j}$ is a vector from an ion in the $i$ sublattice to a nearest neighbor in the $j$ sublattice, and the sum is over all such vectors related by symmetry. The eigenvalue problem can be solved in terms of the real space exchange constants $J_{\alpha\beta}$ calculated from the total energies of collinear magnetic structures.



To calculate the total energy  we use DFT as implemented in the Vienna ab initio simulation package (VASP.5.3)~\cite{kresse1993ab,kresse1996efficient}. The electronic structure is described in the local density approximation (LDA) and the generalized gradient approximation (GGA). {Projector augmented wave (PAW) pseudopotentials \cite{blochl1994projector} with the Perdew-Wang 91 gradient-corrected functional are used. A 500 eV plane-wave cutoff and a $6\times6\times6$ Monkhorst-Pack $k$-point mesh was found to lead to well converged results. We use the atomic positions from the experimental structural parameters (Tab.~\ref{tab:stru}) \cite{Geller1957,Geller1959}. 

\begin{table}[h]
\begin{ruledtabular}
\begin{center}
\begin{tabular}{ccccc}
         & Wyckoff Position  & $x$ & $y$ & $z$ \\
\hline
     Fe$^{\mathrm{O}}$  & 16a  & 0.0000 & 0.0000 &0.0000 \\
     Fe$^{\mathrm{T}}$  & 24d  & 0.3750 & 0.0000 & 0.2500 \\
     Y & 24c  & 0.1250 & 0.0000 & 0.2500 \\
     O & 96h  & 0.9726 & 0.0572 & 0.1492 \\
\end{tabular}
\caption{Atomic positions in the YIG unit cell. The lattice constant is $a=12.376$~\AA.}
\label{tab:stru}
\end{center} 
\end{ruledtabular}
\end{table}


For the (ferrimagnetic) ground-state  structure, the calculated spin magnetic moment of the Fe ions and the electronic band gap of YIG are shown in Fig.~\ref{band}(a). The total moment (including Fe, Y and O ions) per formula unit is consistently 5 $\mu_B$, in good agreement with experimental data \cite{Baettig2008,Pascard1984}. The majority of the moment within the unit cell is highly localised to the Fe sites. In the DFT-LDA calculation, the spin moments are -3.49$\mu_{\rm B}$ for Fe$^{\rm O}$, 3.47$\mu_{\rm B}$ for Fe$^{\rm T}$, and the electronic band gap has the value 0.35~eV, much lower than the value of 2.85~eV found experimentally \cite{Metselaar1974,Wittekoek1975}. Density-functional theory in its bare form is not good at
predicting the energy gap of insulators. This can be overcome to some extent by the inclusion of an on-site Coulomb correction (LDA/GGA+$U$). In this study the Hubbard $U$ and Hund's $J$ parameters for the Fe atoms are determined~\cite{Ching2001,Rogalev2009,Jia2011} by DFT-GGA+$U$ calculations with $U - J$ in the range 0.7 $\sim$ 5.7 eV. The electronic energy gap as well as the spin moments increases slightly with $U - J$. Even for the largest values of $U-J$, the moments are much smaller than expected for pure Fe$^{3+}$ $S=5/2$ state ($\mu_{s} = g\sqrt{S(S+1)} = 5.916\mu_B$), but quite close to those found from neutron diffraction \cite{Rodic1999}. However, these authors suggest that the true space group of YIG is $R\bar{3}$. Only when they perform the refinement in this setting do they obtain good agreement with the known net moment of YIG. The moments obtained are very similar to those found here and by other ab initio calculations (Table \ref{tab:magmom}).  The difference between the $Ia\bar{3}d$ and $R\bar{3}$ groups appears to be sufficiently small to not affect the results much. The electronic energy gap is still smaller than the experimental value, but an even larger $U-J$ causes unwanted artifacts such as a negative gap for spin-flip excitations.

\begin{table}
\begin{ruledtabular}
\begin{tabular}{c c c l l}
\multicolumn{3}{c}{$\mu_{s}$ ($\mu_{\mathrm{B}}$)}  \\
$\mathrm{Fe^{T}}$ & $\mathrm{Fe^{O}}$ & per formula unit & Method & Source
\\
\hline
5.37 & 4.11 & 7.89 & neutron ($Ia\bar{3}d$) & Ref.~\onlinecite{Rodic1999} \\
4.01 & 3.95 & 4.13 & neutron ($R\bar{3}$)\footnote{Fe sites in the $R\bar{3}$  space group retain the tetrahedral and octahedral coordinations. } &  \\
1.56 & 0.62 & 3.44 & LSDA & Ref.~\onlinecite{Xu2000} \\
3.36 & 3.41 & 3.26 & LDA & Ref.~\onlinecite{Baettig2008} \\
3.95 & 4.06 & 3.73 & GGA$+C$ & Ref.~\onlinecite{Jia2011} \\
\hline
3.47 & 3.49& 3.43& LDA & this work\\
4.02 & 4.12& 3.82& GGA$+U$& ($3.7$ eV)  \\

\end{tabular}
\caption{Comparison of magnetic moments in the literature. Note that ‘per formula unit’ includes only the Fe moments and not the total moment of the unit cell. All ab initio studies are for the $Ia\bar{3}d$ point group.\label{tab:magmom}}
\end{ruledtabular}
\end{table}

\section{EXCHANGE INTERACTIONS}

\subsection{Nearest Neighbour}

\begin{figure}
        \centering
        \subfigure[]{
                \scalebox{0.35}[0.35]{\includegraphics[140,00][700,660]{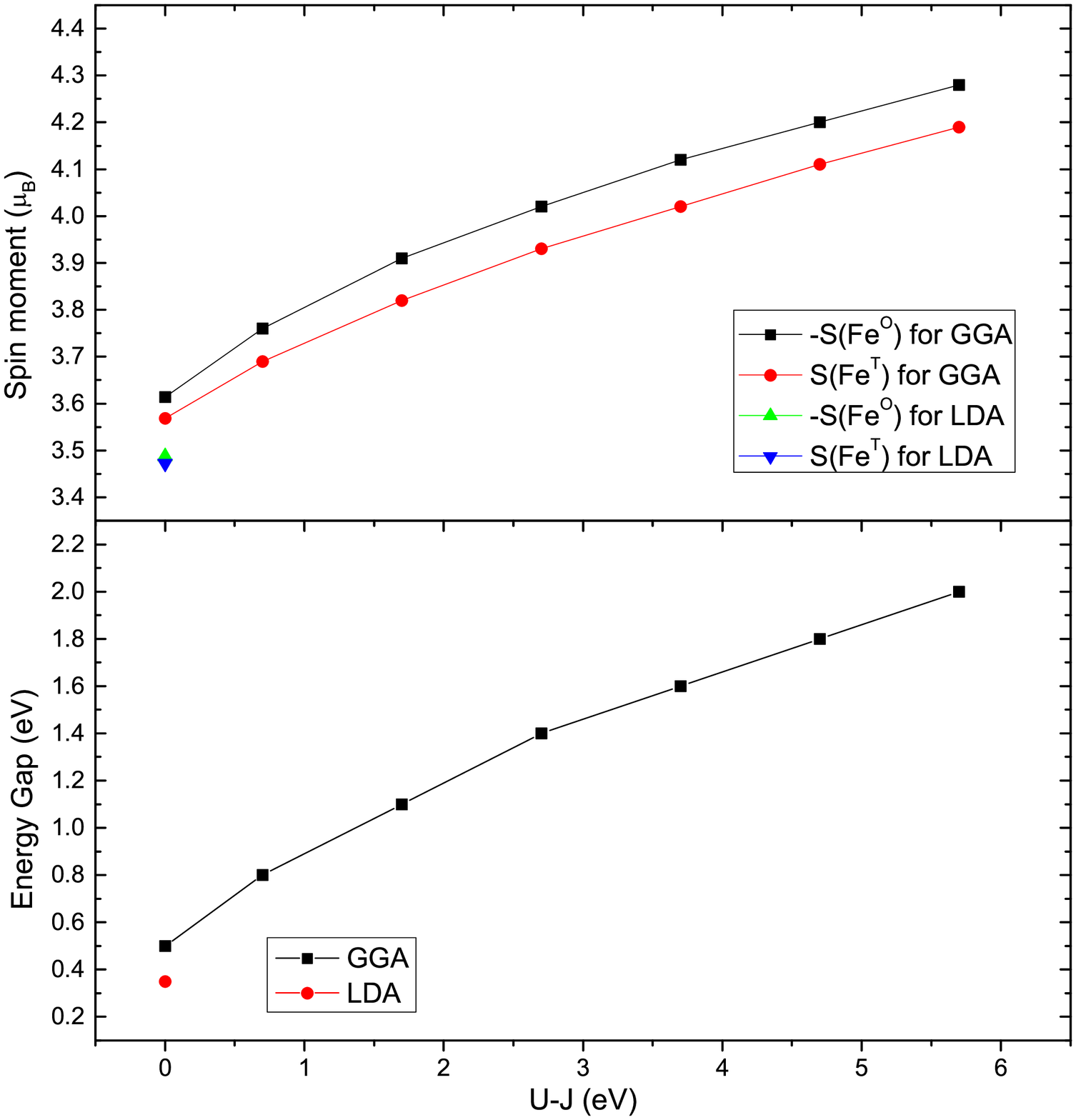}}
        }
        \subfigure[]{
                \scalebox{0.2}[0.23]{\includegraphics[100,60][620,550]{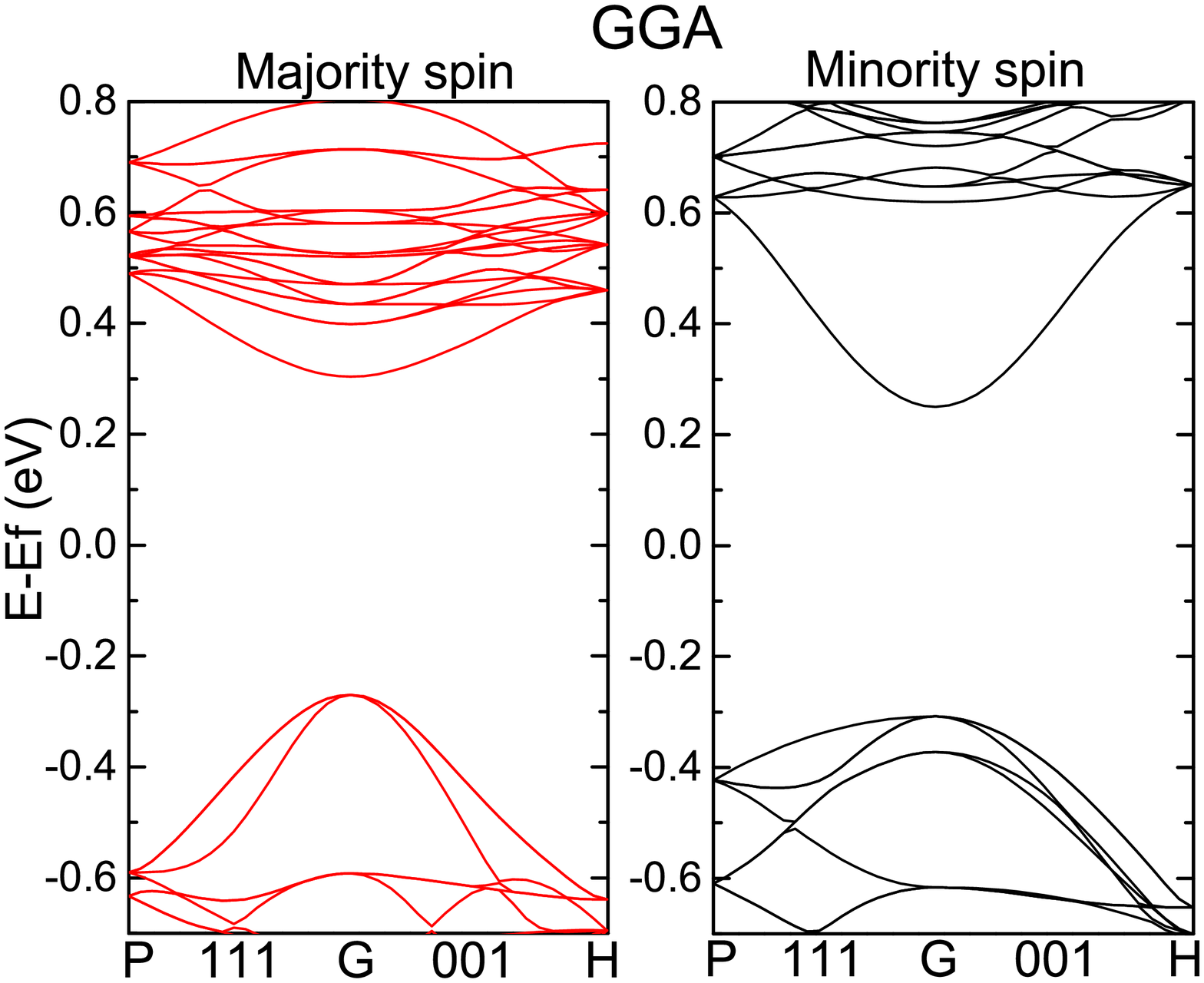}}
                \scalebox{0.2}[0.23]{\includegraphics[0,60][600,550]{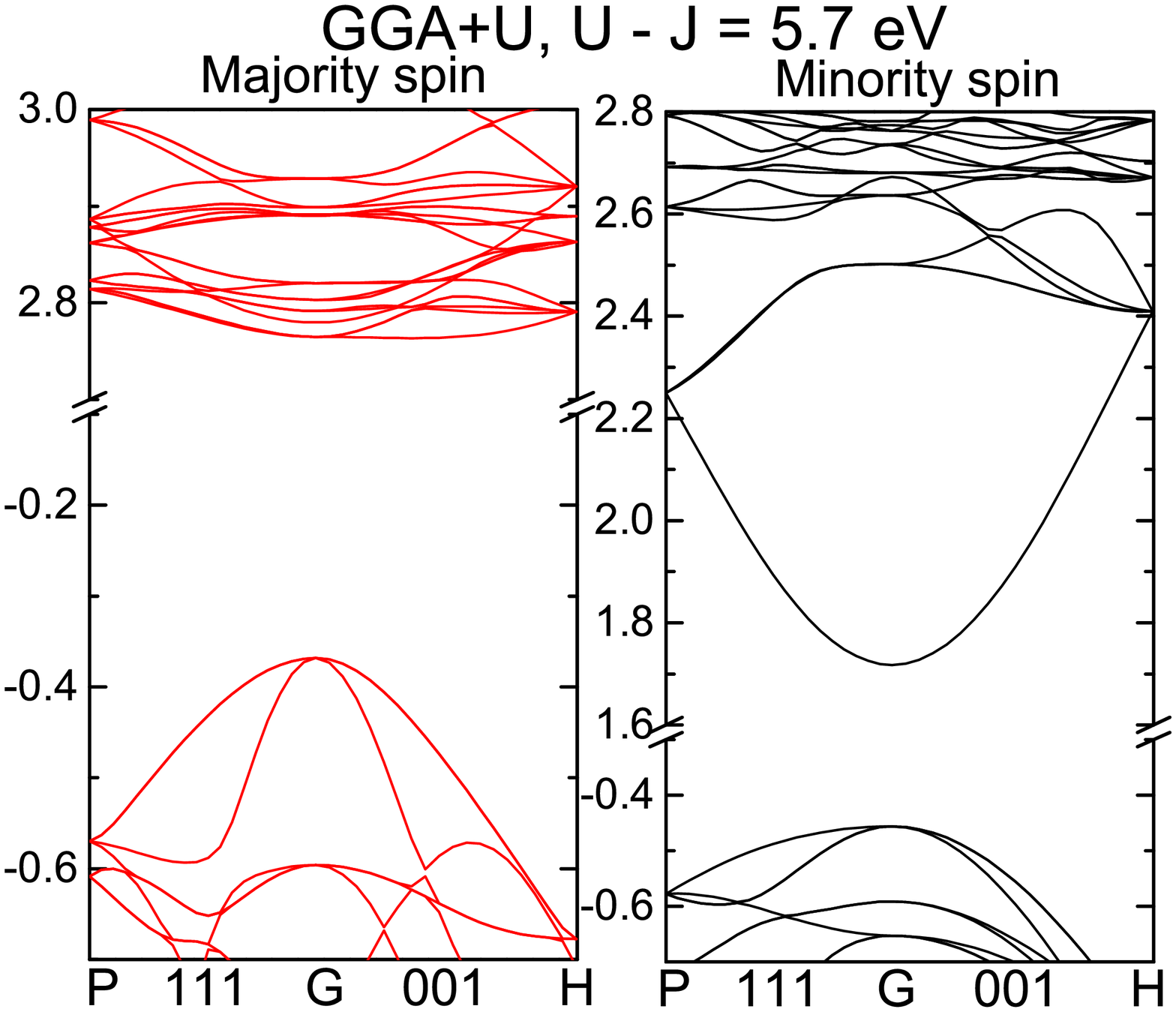}}
        }
        \caption{(a) Spin moments of Fe ions (per panel) and band gap of YIG (lower panel) obtained by computed in the LDA, GGA, and GGA+$U$ approximations. Symbols mark calculated values and solid lines are guides for the eye. (b) The band structures of YIG in the GGA (left) and GGA+$U$, $U - J=5.7$ ${\rm eV}$ (right) calculations. \label{band}}
\end{figure}


Ten different spin configurations (SC) were used to determine the exchange constants. Considering the NN model first, with $E_{aa}=J_{aa}S_aS_a$, $E_{dd}=J_{dd}S_dS_d$ and $E_{ad}=J_{ad}S_aS_d$, where $S_a$, $S_d$ are the +/- directions of Fe$^{\rm O}$, Fe$^{\rm T}$ ions, the total energies Eq.~(\ref{eq:energy}) are listed in Tab. \ref{tab:etotnn}. 

\begin{table}[h]
\caption{Total energies for different spin configurations (SC) in the NN model. SC (a) is the ground-state structure. The other configurations are gotten by changing the magnetization directions of part of Fe ions.}
\begin{center}
\begin{tabular}{clcl}
\hline\hline
     SC    & $E_{tot}$   & SC    & $E_{tot}$ \\
\hline
     a     & $E_0+32E_{aa}+24E_{dd}+48E_{ad}$   & f  & $E_0+32E_{aa}-24E_{dd}$ \\
     b     & $E_0+32E_{aa}+24E_{dd}-48E_{ad}$   & g  & $E_0-32E_{aa}-24E_{dd}$ \\
     c     & $E_0+32E_{aa}+8E_{dd}+32E_{ad}$    & h  & $E_0-32E_{aa}-8E_{dd}$  \\
     d     & $E_0+32E_{aa}-8E_{dd}+16E_{ad}$    & i  & $E_0-32E_{aa}+8E_{dd}$  \\
     e     & $E_0+16E_{aa}+16E_{dd}+28E_{ad}$   & j  & $E_0-32E_{aa}+24E_{dd}$ \\            
\hline\hline
\end{tabular}
\end{center} \label{tab:etotnn}
\end{table}
        
The exchange constants are the solutions of each of four linear equations. To minimize the dependence of the results on the choice of the spin configurations, the final results were obtained using all the configurations (a)-(j) listed in Tab. \ref{tab:etotnn}. The final values, shown in Fig.~ \ref{fignnj}, were obtained by a least squares fit of the 10 SC's. In the DFT-LDA/GGA calculations, the exchange constant $J_{dd}$ is negative, meaning that this interaction favors ferromagnetic order. This result contradicts all previous results in the literature~\cite{Strenzwilk1968,Cherepanov1993} - indicating that the DFT-LDA/GGA method fails to describe the magnetism of YIG. However, in the GGA+$U$ method, all three exchange constants are positive (antiferromagnetic), $J_{dd}$ is an order of magnitude smaller than $J_{ad}$, while $J_{aa}$ is about half of $J_{dd}$. The strong $J_{ad}$ inter-sublattice exchange dominates the smaller intra-sublattice energies, forcing the ferrimagnetic ground state of the bulk. All the three exchange constants decrease as $U - J$ increases, because a larger on-site $U - J$ of the Fe atoms leads to a more localized electronic structure resulting in weaker exchange. Previous works assumed that $J_{ad} \gg J_{aa}, J_{dd}$, which is required to constrain the fitting problem~\cite{Harris1963,Strenzwilk1968,Plant1983,Cherepanov1993}. Our results show directly the smallness of the intra-sublattice exchange energies because of a stronger objective function  for the least-squares fitting procedure.

\subsection{Next Nearest Neighbour}

\begin{figure}
       \scalebox{0.37}[0.35]{\includegraphics[147,54][705,532]{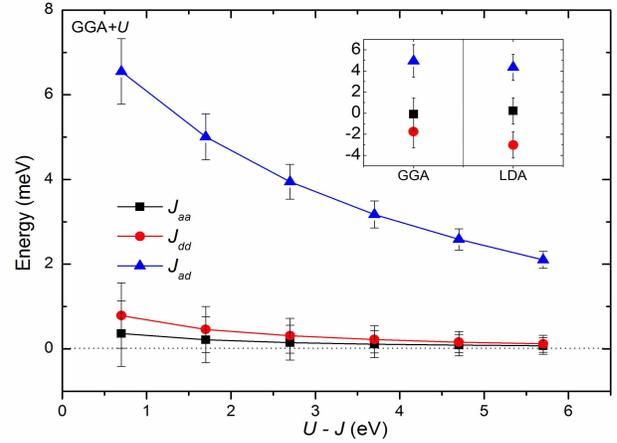}}
        \caption{Calculated exchange constants (in units of meV) by the DFT-GGA+$U$ method. The error bars denote the square root of the squared 2-norm of the residual ($l^2$-norm). Exchange constants favoring a ferromagnetic alignment are here denoted negative. (Insert) Calculated exchange constants (in units of meV) in the DFT-GGA/LDA approximations. \label{fignnj}}
\end{figure}

The error bars in Fig.~\ref{fignnj} reveal a large covariance in  the fitting of the NN spin model to the different configurations.  Even though the errors decrease with increasing $U - J$, the variance in the energies is still comparable to its estimation. This situation can be improved by extending the NN to the NNN model with additional parameters $J'_{aa}$, $J'_{dd}$ and $J'_{ad}$. The total energies of the corresponding SC can be rewritten (shown in Tab.~\ref{tab:etotn3}), where $E'_{aa}=J'_{aa}S_aS_a$, $E'_{dd}=J'_{dd}S_dS_d$, $E'_{ad}=J'_{ad}S_aS_d$ and $E_{tot}$ stands for the total energy expression in the NN model. The exchange constants are obtained from the set of linear equations for the SC (a)-(g) listed in the table. SC (h)-(j) are selected to check whether the results are reasonable.
$E_{cal}$ are the calculated total energies for $U - J = 4.7$ ${\rm eV}$ relative to the ground state (SC (a)). The energy difference for the different SC is of the order of $1 \sim 10$ eV which is much larger than the accuracy of the calculation ($10^{-3}$ eV). $\Delta E_{\rm NNN}$ ($\ll \Delta E_{\rm NN}$) is the difference between the total energies calculated ab initio and the fitted total energies from the NNN (NN) spin model and constitutes the energy that has not been accounted for in our model Hamiltonian. This can be, for example, from longer ranged exchange interactions or anisotropies in the system. The difference between the first-principles total energy and the spin model $|\Delta E_{\rm NN}|$ amounts to up to $7.85\%$, but the NNN model has a significantly smaller value $|\Delta E_{\rm NNN}| = 0.66\%$, which we deem to be acceptable.

In table~\ref{tab:exchange} we compare our results to other values in the literature. Almost all of the exchange interactions we calculated are lower than obtained from fitting experimental data. Especially the $J_{ad}$, the strongest interactions, is lower than others have suggested, although the NNN $U-J = 3.7$ eV is quite close. Lowering $U-J$ gives an increase in $J_{ad}$, but at the expense of the size of the magnetic moments and the width of the electronic band gap. One may naively think that lower exchange constants will give a lower Curie temperature, however because the intra-sublattice interactions are also antiferromagnetic in character the situation is more complicated.

Where NNN values are calculated  the order of magnitude agrees with attempts by Plant to fit the neutron scattering data with a NNN model~\cite{Plant1983}.
\begingroup
\squeezetable 
\begin{table}[h]
\begin{ruledtabular}
        \caption{Total energies for different SC in the NNN model. The energies are in units of meV.  $E_{tot}$ and  $E'_{tot}$ are the total energies for the NN and the NNN models. $E_{cal}$ are the total energies calculated ab initio and $\Delta E_{\rm NNN}$ ($\Delta E_{\rm NN}$) are the differences between the fitted total energies from the NNN (NN) spin model and $E_{cal}$. $E_{cal}$ of the ground-state structure (SC (a)) is denoted zero.}
        \begin{center}
                \begin{tabular}{ccccc}
                        SC    & $E'_{tot}$   & $E_{cal}$    & $\Delta E_{\rm NNN}$   & $\Delta E_{\rm NN}$\\
                        \hline
                        a     & $E_{tot}+24E'_{aa}+48E'_{dd}+48E'_{ad}$   &    0.00 & \ 0.37 & -59.97\\
                        b     & $E_{tot}+24E'_{aa}+48E'_{dd}-48E'_{ad}$   & 4225.32 &  -0.31 &  -3.69\\
                        c     & $E_{tot}+24E'_{aa}+16E'_{dd}+32E'_{ad}$   & 1907.02 & \ 0.39 & -58.19\\
                        d     & $E_{tot}+24E'_{aa}+16E'_{dd}+16E'_{ad}$   &  566.01 & \ 0.38 &\ 44.42\\ 
                        e     & $E_{tot}+12E'_{aa}+32E'_{dd}+32E'_{ad}$   &  778.86 & \ 0.23 & \ 5.97\\ 
                        f     & $E_{tot}+24E'_{aa}+48E'_{dd}$          & 1987.42 &  -0.21 & -36.19\\ 
                        g     & $E_{tot}+24E'_{aa}+48E'_{dd}$          & 1228.54 & \ 0.24 &\ 52.29\\
                        h     & $E_{tot}+24E'_{aa}+16E'_{dd}$          & 1848.59 &  -3.04 &\ 43.44\\
                        i     & $E_{tot}+24E'_{aa}+16E'_{dd}$          & 1885.68 &  -7.62 &\ 49.55\\
                        j     & $E_{tot}+24E'_{aa}+48E'_{dd}$          & 2018.23 & -13.40 & -39.80\\
                \end{tabular}
        \end{center} \label{tab:etotn3}
        \end{ruledtabular}
\end{table}
\endgroup

Compared with the NN model (as shown in Tab.\ref{tab:exchange}), the values of $J_{aa}$, $J_{dd}$ and $J_{ad}$ in the NNN model became slightly smaller but still obey $J_{ad} \gg J_{dd} > J_{aa}$. The additional interactions $J'_{dd}$ and $J'_{ad}$ are of the same order of magnitude as the NN intra-sublattice exchange and are also antiferromagnetic. Notably, $J'_{dd} > J_{dd}$ interaction.
 
\begin{table*}[t]
\begin{ruledtabular}
\begin{tabular}{c c c c c c l l}
\multicolumn{6}{c}{(meV)}  \\
$J_{ad}$ & $J_{dd}$ & $J_{aa}$ & $J'_{ad}$ & $J'_{dd}$ & $J'_{aa}$ & method
& reference \\
\hline
3.10 & 1.40 & 0.96 & - & - & - & molecular field approximation & Ref.~\onlinecite{Anderson1964}\\
3.90 & 0.78 & 0.78 & - & - & - & magnetization fit & Ref.~\onlinecite{Harris1963}\\
3.40 & 0.69 & 0.69 & - & - & - & neutron spectrum fit* & Ref.~\onlinecite{Plant1977}\\
2.60 & 1.00 & 0.56 & - & - & - & molecular field approximation & Ref.~\onlinecite{Srivastava1982}\\
3.20 & 0.45 & 0.00 & 0.23 & 0.14 & 0.75 &  neutron spectrum fit*& Ref.~\onlinecite{Plant1983}\\

3.40 & 1.20 & 0.33 & - & - & - &  neutron spectrum fit* & Ref.~\onlinecite{Cherepanov1993}\\
\hline
3.176 & 0.223 & 0.112 & - & - & - & ab initio GGA$+U$ ($U-J=3.7$~eV) & this
work\\
2.917 & 0.213 & 0.090 & 0.218 & 0.228 & 0.005 & &\\
2.584 & 0.160 & 0.091 & - & - & - & ab initio GGA$+U$ ($U-J=4.7$~eV) & \\
2.387 & 0.154 & 0.072 & 0.163 & 0.179 & 0.004 & &\\
\end{tabular}
\end{ruledtabular}
\caption{Comparison of exchange constants in the literature. (*) all fits
to neutron data use the same data from Plant~\cite{Plant1977}.\label{tab:exchange}}
\end{table*}


\begin{figure}
        \centering
        \includegraphics[width=0.47\textwidth]{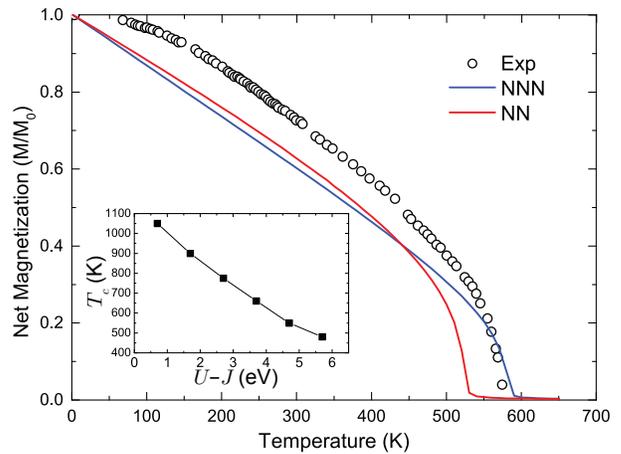}
        \caption{The magnetization curves of the NN model (red line) and the NNN model (blue line) with exchange constants fitted to the ab initio energies for $U - J = 4.7 $ ${\rm eV}$ for the NN model and $U - J = 3.7 $ ${\rm eV}$ for the NNN model. The experimental data \cite{Anderson1964} are indicated by circles. (Insert) The Curie temperatures of the NN model fitted to the ab initio results for different $U - J$.\label{fig5}}
\end{figure}

\section{Intrinsic Properties}
\subsection{Curie Temperature and Magnetization}
We calculate the temperature dependence of the magnetization and the Curie temperature ($T_C$) from the spin models by Metropolis Monte Carlo (MC) simulations on a 32 $\times$ 32 $\times$ 32 super cell (each unit cell contains 20 spins) with periodic boundary conditions \footnote{To ensure thermal equilibrium, the convergence of the magnetization was subjected to a Geweke diagnostic test~\cite{Geweke1991}. The final $80\%$ of the data was  used to calculate the thermally averaged magnetization.}. The temperature dependence of the total magnetization, $M = M_{d} - M_{a}$, is shown in Fig.~\ref{fig5}, normalized by $M(T=0~\mathrm{K})$. The $T_C$ of the NN model exchange parameters using different $U - J$ values are shown in the inset. The experimental value of $T_C$ is 570~K \cite{Anderson1964,Nimbore2006}. In the NN model, the larger $U$ gives smaller exchange constants and hence weaker interactions giving a lower $T_C$. This follows intuitively because of the increased localisation of the wave functions reducing the exchange and hence also the Curie temperature. With the parameters $U-J=4.7$ ${\rm eV}$, $T_C$ is 540 K, in good agreement with the experimental value. The magnetization curve of the NNN model is quite similar to the NN model with a slightly higher $T_C$ of 590 K using the parameters exchange parameters when $U-J=3.7$ ${\rm eV}$.} The finite slope at low temperatures in both models does not agree with experiments. This deviation is ascribed to our disregard of quantum statistics in the simulations. Nevertheless, at higher temperatures the calculated shapes of the magnetization and  $T_C$  agree well with experiments.

\subsection{Spin wave spectrum}

\begin{figure}
	\includegraphics[width=0.43\textwidth]{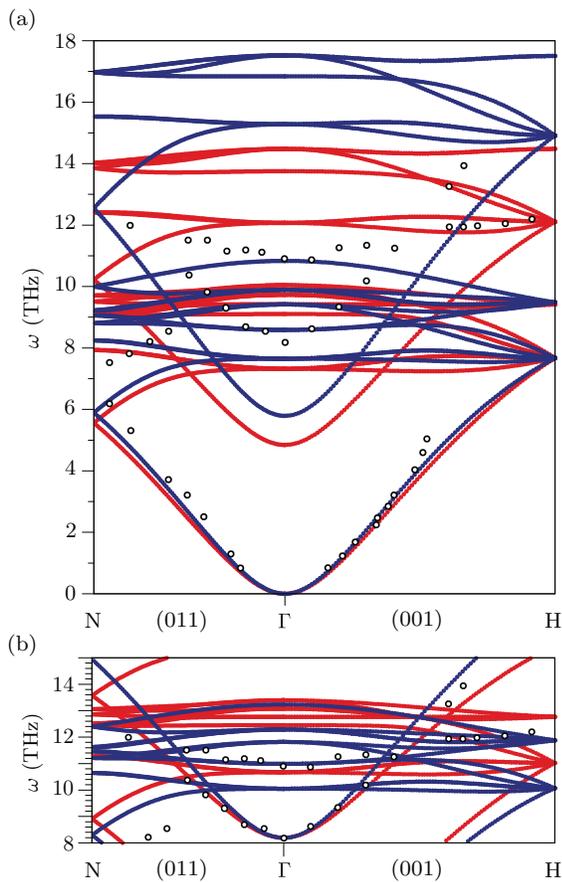}
	
	\caption{Spin-wave spectrum in the first Brillouin zone for the NN model (red dots) derived from ab initio calculations with $U-J=4.7$ ${\rm eV}$ and the NNN model (blue dots) where $U-J=3.7$ ${\rm eV}$ and compared to the available neutron scattering data (black circles) \cite{Plant1977}. (a) The entire spin wave spectrum. (b) Comparison of the shape of the parabolic optical mode the results are  shifted by +3.35 THz for the NN model and +2.40 THz for the NNN model and compared to the 83~K experimental data. The directions in $k$-space use the standard labels of bcc reciprocal lattice.\label{fig4}}
\end{figure}

Next we calculate the spin wave spectrum from our parameterized Heisenberg model. We choose the exchange constants with the parameter $U-J=4.7$ ${\rm eV}$ for the NN model and the parameter $U-J=3.7$ ${\rm eV}$ for the NNN model. The analytic results of the spin-wave spectrum Eq.~(\ref{rmat}) are shown in Fig.~\ref{fig4}. The experimental data from Refs.~\onlinecite{Plant1977,Plant1983} are for 83~K. Strictly speaking only the low temperature results should be compared with theory. 

\emph{Dispersion relation of the acoustic mode} -- The slopes of the lowest acoustic mode of the NN model and the NNN model both agree well with the neutron scattering data (Fig.~\ref{fig4}(a)). The spin-wave stiffness $D$ is governed by the second derivative at the $\Gamma$-point. $D=77 \times 10^{-41} ~\rm J \cdot \rm m^2$ and $85 \times 10^{-41}~\rm J \cdot \rm m^2$ for the NN and 
NNN models, respectively. The values reported in the literature obtained by different experimental methods \cite{Anderson1964,Cherepanov1993,Srivastava1987} vary from $D = 42 \times 10^{-41}~\rm J \cdot \rm m^2$ to $109 \times 10^{-41}~\rm J \cdot \rm m^2$.

\emph{High frequency modes} -- As shown in Fig.~\ref{fig4}(a), the spectra of both models in the range of 8 THz $\sim$ 11 THz have a similar structure. However, the modes of the NNN model are more separated, especially at the $\Gamma$-point, which we ascribe to $J'_{dd}$. At high frequencies (above 12 THz), the modes of the NNN model have much higher frequency compared to the corresponding ones of the NN model.

\emph{Spin wave gap} -- The (exchange) gap between two lowest (acoustic and optical) modes at the $\Gamma$-point of the NN model is about 5 THz, while the one of the NNN model is 0.945 THz higher due to the larger $J_{ad}$ in the latter, but is still smaller than the experimental gap of about 8 THz at 83~K. The comparison of the frequency-shifted second lowest mode with the experimental data are shown in Fig.~\ref{fig4}(b). The slope of the NNN model is a little steeper than that of the one of the NN model, and they are both in good agreement with the experimental data. 


In conclusion, we report exchange constants of YIG computed from first principles but with an adjustable $U-J$ constant to increase the density functional  band gap. We found that NNN interactions are required for a good fit of total energies by a Heisenberg model. Our results reproduce the experimental Curie temperature well. In addition, we obtain a spin-wave spectrum in which the lowest acoustic mode agrees very well with the available neutron scattering data. However the lowest optical mode energy appears to be underestimated, emphasizing the need for more studies of the temperature dependent spin wave spectrum.

\begin{acknowledgments}
This work was supported by the National Natural Science Foundation
of China (Grants No. 61376105, No. 21421003 and No. 11374275) and JSPS KAKENHI Grant Nos. 25247056, 25220910, 26103006. JB acknowledges support from the Graduate Program in Spintronics, Tohoku University. LSX and JB acknowledge support from the JST Sakura Science Exchange Program.

\end{acknowledgments}

\bibliography{yig_abinitio_fixed}

\end{document}